\documentclass[prb,twocolumn,aps,showpacs]{revtex4}
\makeatletter
\def\@pnumwidth{2em}
\makeatother
\usepackage{graphicx}
\usepackage{psfrag}
\usepackage{dcolumn}
\usepackage{bm}
\def\ds{\displaystyle}
\def\be{\begin{equation}}
\def\ee{\end{equation}}
\def\im{I_{\rm min}}
\def\eps{\epsilon}
\begin{document}

\title{Character of eigenstates of the three- dimensional  disordered Hamiltonian}

\author{J. Brndiar$^1$ and P. Marko\v{s}$^{1,2}$}%
\affiliation{$^1$Institute of Physics, Slovak Academy of
Sciences,  845 11 Bratislava, Slovakia\\
$^2$Dept. Physics,  Faculty of Electrical Engineering and Information Technology, 
Slovak University of Technology, Ilkovi\v cova 3, 812 19 Bratislava, Slovakia
}

\begin{abstract}
We study numerically the character of electron eigenstates of the three dimensional disordered 
Anderson model.  Analysis of the statistics of inverse participation ratio as well 
as numerical evaluation of the electron-hole correlation function  confirm that there are no 
localized states below the mobility edge,  as well as no metallic state in the tail 
of the conductive band. 
We discuss  also finite size effects  observed in the analysis of all the 
discussed quantities.
\end{abstract}

\pacs{73.23.-b, 71.30., 72.10. -d}

\maketitle

\section{Introduction}

Localization of electron in disordered systems, \cite{Anderson-58}
manifests the wave character of the electron propagation. Components of the wave function,
scattered on randomly distributed impurities,
interfere with each other. This interference might lead to   the electron localization.

While all states are localized in one dimensional systems, 
the localization of all electronic states in the three dimensional (3D) systems
appears only when 
the strength of the disorder, $W$,   exceeds certain critical value $W_c$.
For weaker disorder, $W<W_c$, the energy band is divided into two parts, 
separated by the mobility edge $E_c$. It is supposed that all states 
with energy $|E|<E_c$ are metallic (conductive), while  only
localized states exist for  $|E|>E_c$.

Although electron localization is easy to understand intuitively,
the wave character of electron propagation causes new, nontrivial phenomena in all three transport regimes.

In the limit of weak disorder, the system is metal, but the 
scattering of  electron on  impurities is responsible for non-classical 
phenomena, 
such as the universal, system size independent
conductance fluctuations \cite{LSF,PNato} and weak localization (antilocalization). \cite{bergman,Kramer}
The complete description of the transport was given by Green's function
method, \cite{LSF} random matrix theory, \cite{PNato,been} DMPK equation \cite{DMPK} or
supersymmetric methods. \cite{mirlin} 

In the opposite limit of strong disorder, (localized regime), 
the fluctuations of the conductance are so strong, 
(they exceed the  mean conductance  in many orders of
magnitude),  that the conductance itself is not a relevant parameter of the
theory any more. Instead,
the logarithm of the 
conductance must be studied. \cite{ATAF} 

Owing to huge fluctuations, 
the analytical description  of the critical regime 
is much more complicated.
Thus, although the critical metal-insulator
transition is well understood by the single parameter scaling,
\cite{AALR}  quantitative estimation
of critical parameters is still almost unsolvable problem.
Of particular interest is the 
critical exponent $\nu$ \cite{wegner}  which determines the divergence of the
correlation length, $\xi\propto |E-E_c|^{-\nu}$  at the  mobility edge.
Over two decades, there is a discrepancy between 
theoretical predictions and numerical estimations  of $\nu$.
For 3D Anderson model,  numerical results $\nu\approx 1.5 - 1.57$
obtained by various numerical methods \cite{MacKK,SO,SMO1,shklovski,isa1995},
overestimate the analytical prediction, $\nu=1$ \cite{VW,suslov}.
The disagreement is even worse in 4D (numerical data
\cite{travenec,acta,cuevas} give
 $\nu\approx 1$, while theory predicts $\nu=0.5$).  
Numerical data for $\nu$ in low dimensional systems
$d=2+\epsilon$ ($\varepsilon\ll 1$) 
\cite{travenec,acta} also  do not agree
with the analytical $\varepsilon$ - expansions. \cite{wegner} 

In our opinion, the origin of this discrepancy lies 
in the procedure of the averaging over the disorder. 
\cite{ATAF,acta} While averaging is extremely difficult to perform 
analytically  it is straightforward in numerical simulations.
Therefore, we expect that numerically obtained critical parameters will be,
soon or later, supported by analytical results. 
\cite{kawabata,AGG}

The inhomogeneous spatial distribution of electrons
in the critical and localized regimes \cite{janssen,EM}
inspire people to build mean field theories on the 
the analysis of the statistics of the local density of state. 
\cite{dobro,song}  
Recent numerical data \cite{Markos,prior} led to 
new analytical theories of the transport in the strongly localized regime. 
\cite{mukl,MMWK,somoza}

In this paper we study numerically the character of eigenstates of the
disordered 3D Anderson Hamiltonian. 
Our aim is to exclude any possibility for the existence of localized states below the
mobility edge, $E_c$. 
The performed  analysis is inspired
 by recent analytical theory of the Anderson transition \cite{JK}
which predicts that the number of metallic states decreases continuously to zero when Fermi
energy approaches the mobility edge from the metallic side. The idea is formulated 
in terms of the electron-hole correlation function $\Gamma_q(E,\omega)$ (defined later
by Eqs. \ref{propagator} and \ref{propagator-q})
which possesses
in the limit of small energy difference $\omega$ and small wave vector $q$
the diffusive pole of the form
\be\label{jeden}
\Gamma_q(E,\omega)=\ds{\frac{2\pi\rho(E)}{-iA(E)\omega+D(\omega)q^2}}.
\ee
($\rho(E)= {\rm Im}~G(E+i\eps)/\pi$ is the density of states \cite{economou} determined 
by the one electron Green function $G(E+i\eps)$, and $D$ is the diffusion constant).

Expression (\ref{jeden}) differs from the ``classical'' diffusion pole 
\cite{VW} by the presence of
the function $A(E)$. It is claimed in Ref. \cite{JK} that
$A(E)$   increases when $E$ approaches the mobility edge
 and becomes infinite at the critical point. The ratio $\rho(E)/A(E)$ determines 
the portion extended (diffusive) states from all available states, given by $\rho(E)$.
The rest states, $\rho(E)\times(A-1)/A$, are  spatially localized, 
although $E$ lies in  the metallic phase, $|E|<E_c$. 

Intuitively, the existence of localized states in the metallic phase 
seems  to be  impossible 
\cite{Kramer}. It also contradicts analytical analysis of the electron eigenstates
\cite{mirlin,fyod,Falko}. Nevertheless,  no numerical analysis of this problem
has been performed yet.  
The present paper fills this gap.

We investigate in Section II
the singular behavior of   $\Gamma_{q}(E,\omega)\sim\omega^{-1}$ for $q=0$ and  prove that 
$A(E)\equiv 1$ for all energies $E$, both in the metallic and localized regime.
This confirms  theoretical expectations \cite{VW, EM}.
We also analyze the  diffusive pole in the metallic phase
(band center), find diffusive constant
and discuss statistical properties of the function $\Gamma_q(E,\omega)$.

Another proof of the absence of localized states in the metallic phase is given in 
Section III, where we 
study the probability distribution of the  inverse participation ratio (IPR) \cite{Kramer}
defined later by Eq. \ref{ipr}.
Statistical properties of IPR were analytically studied in \cite{mirlin,fyod,Falko}.
Statistical properties
of IPR at the critical point were subject of analytical and numerical analysis
in connection to the multifractal spatial distribution of electrons.
\cite{janssen,EM,evers2000} 
Scaling of IPR in the critical region was proved in \cite{BM}. 
Here, we discuss how the probability distribution of IPR depends on the 
system size and the distance $E-E_c$ of the energy from the mobility edge. Our data show that
the probability to find the localized state in the metallic phase decreases
exponentially when the size of the system  increases.

Electron eigenenergies and wave functions are calculated for the 3D
Anderson Hamiltonian,
 \be\label{ham}
 {\cal{H}}=\sum_{\vec{r}} \epsilon_{\vec{r}} c_{\vec{r}}^\dag c_{\vec{r}}+V\sum_{[\vec{r}\vec{r'}]}c_{\vec{r}}^\dag c_{\vec{r'}}.
 \ee
 Here,  $\vec{r}$ determines the  site in  the 3D lattice of the size $L^3$,
 $\epsilon_{\vec{r}}$ is the random energy
 distributed with 
the Gaussian distribution, $P_G(\epsilon_{\vec{r}})=(2\pi W^2)^{-1/2}\exp (-\epsilon_{\vec{r}}^2/2W^2)$.
 Parameter $W$ measures the strength of the disorder and $V=1$ determines the energy scale.
 For $E=0$, the critical disorder
 $W_c\approx 6.15$. We fix the strength of the disorder $W=2$ throughout this paper.
Then, the mobility edge $E_c=6.58$  separates the metallic and insulating phase.
\cite{BM}

\section{The electron-hole correlation function}

In this section we investigate
 the electron-hole correlation function defined as
\be\label{propagator}
\Gamma(E,\omega,\vec{r'},\vec{r})
=\langle G(E+\omega/2+i\epsilon,\vec{r},\vec{r'})G(E-\omega/2-i\epsilon,\vec{r'},\vec{r})\rangle.
\ee
Here,
$G(E+i\epsilon) = \left[E+i\epsilon-{\cal H}\right]^{-1}$
is the one-particle Green's function \cite{economou}, 
which determines the density of states,
and $\langle\dots\rangle$ means averaging over realization of the disorder.

We calculate  the Fourier transformation,
\be\label{propagator-q}
\Gamma_q(E,\omega) = \sum_{\vec{r},\vec{r'}} e^{i\vec{q}.(\vec{r}-\vec{r'})}
\Gamma(E,\omega,\vec{r'},\vec{r}),
\ee
set $q=0$ and analyze the 
singular $\omega$ dependence
\be\label{t}
\Gamma_{0}(E,\omega) = \ds{\frac{B(E)}{-i\omega}},~~~\omega\to 0.
\ee
Comparison with Eq. (\ref{jeden}) gives   $B(E)=2\pi\rho(E)$.
Coefficient $B(E)$ would either equal to  $2\pi\rho(E)$ (case $A\equiv 1$),
or it would decrease to  zero, $B(E)\sim (E_c-E)^{\alpha}$, if the scenario proposed
in Ref. \cite{JK} is true.

\begin{figure}
\psfrag{x}{{${\rm Im}~\Gamma_0$}}
\psfrag{y}{{${\rm P}({\rm Im}~\Gamma_0)$}}
\psfrag{a}{{$\epsilon=5/L^3,\ L=18$}}
\psfrag{e}{{$\epsilon=60/L^3,\ L=18$}}
\psfrag{c}{{$\epsilon=80/L^3,\ L=18$}}
\includegraphics[clip,width=0.45\textwidth]{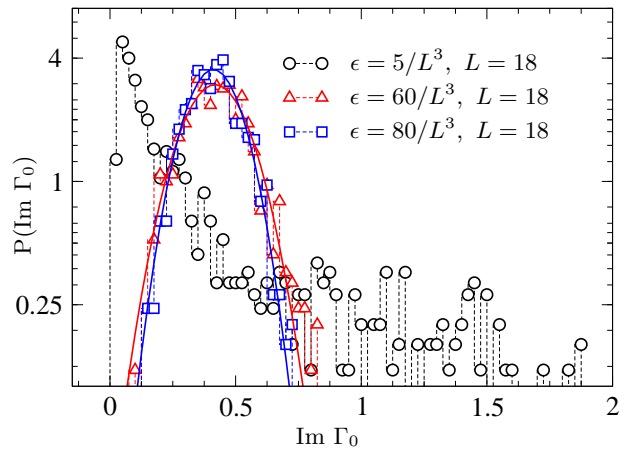}
\caption{(color online)
Probability distribution $P({\rm Im}~\Gamma_0)$ for $E=7$, $\omega=0.14$,  $L=18$
and for three values of  $\epsilon$. The density 
of states, $\rho(E=7)\approx 0.0076$ and the mean level spacing
is $\approx 130/L^3\approx 0.0223$.
The distribution obtained for $\epsilon=5/L^3$ is clearly unphysical,
but the choice $\epsilon=60/L^3$ is already sufficient to reach  the
Gaussian distribution of Im~$\Gamma_0$.
}
\label{distrx}
\end{figure}

\begin{figure}
\psfrag{x}{$1/\omega$}
\psfrag{y}{${\rm Im}~\Gamma_0$}
\includegraphics[clip,width=0.45\textwidth]{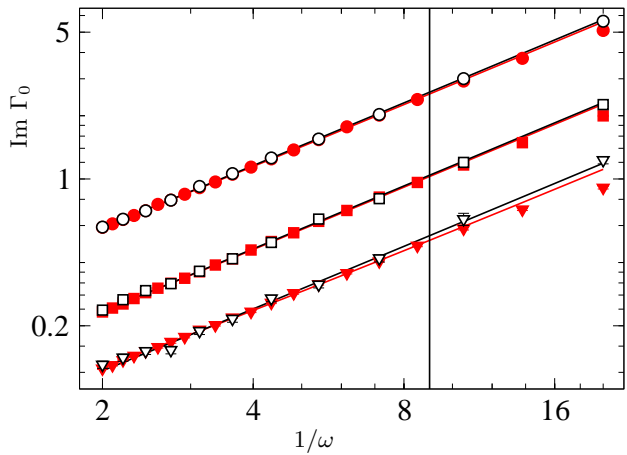}
\caption{(Color online) %
The imaginary part of $\Gamma_0(E,\omega)$, given by Eq. (\ref{t})
as a function of $\omega^{-1}$.
$E=5$, $L=12$ (circles), $E=6.58$, $L=16$  (squares) and $E=7$, $L=18$  (triangles). 
$\eps=5/L^3$ (open symbols). Higher values of $\eps$ were used to check the
stability of data: $\eps=15/L^3$ ($E=6$), $40/L^3$ ($E=6.58$) and $80/L^3$ ($E=7$)
(full symbols).
Only data left of the vertical line were used for calculation of $B(E)$. Solid lines
are fits $\ln$~Im~$\Gamma_0=-\ln\omega + \ln B(E)$.
}
\label{gamma-1}
\end{figure}

\begin{figure}[t]
\psfrag{x}{$2\pi\rho(E)$}
\includegraphics[clip,width=0.45\textwidth]{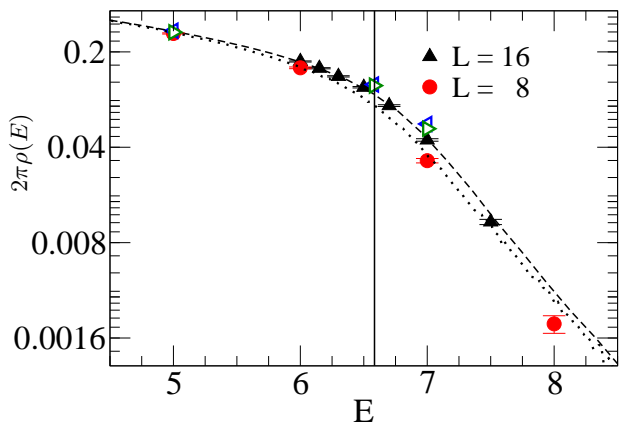}
\caption{(Color online) %
The coefficient $B(E)$  \textsl{vs} energy $E$
compared with the density of states $2\pi\rho(E)$.
Only critical region is shown. Deep in the metallic phase, the coincidence of
$B(E)$ and 
$2\pi\rho(E)$ is even better.
Size of the system is $L=8$ (circles)  and $L=16$ (triangles). 
Open triangles show $B(E)$ calculated in Fig. \ref{gamma-1} for $\eps=5/L^3$ (triangle left)
and for higher values of $\eps$ (triangle right). 
The density of states is calculated for $N_{\rm stat}=1000$ systems  of the size 
$L=16$ (dashed line) and $L=8$ (dotted line).   Dirichlet boundary conditions were used for calculation of both $\rho(E)$ and
$\Gamma_0$. Vertical line marks the position of the critical point.
}
\label{gamma-2}
\end{figure}

Before presenting numerical data, it is worth to comment the numerical method
of calculation  of the Green's functions.
 Our method is based on the numerical inversion
of the matrix  
$E\pm i \epsilon-{\cal H}$.  
For the reliability of data, the choice of the value of the 
small imaginary part of the energy, $\epsilon$,  is crucial.
We expect that $\eps$ should be comparable to  the typical level spacing,
$1/(\rho(E)L^3)$. Using numerical data for the density of state
at the band center, $\rho(E=0)\approx 0.115$,  we choose  
$\epsilon=5/L^3$. This value is sufficiently large to avoid any numerical instabilities
(discussed later in Sect. IIIC) in the band center, but
it  might be too small in the band tail,
where the density of state is much smaller. 
Therefore, various larger values of $\epsilon$ were used to guarantee
the numerical stability of our results.

As an example of how the value of $\epsilon$ influences the accuracy of numerical results,
we show in Fig. \ref{distrx}
the probability distribution $P($Im~$\Gamma_0)$  for energy $E=7$ and the system size $L=18$.
Statistical ensemble of $N=2000$ samples was used to calculate the distribution.
If $\epsilon$ is much smaller than the level spacing,
then the distribution  $P($Im~$\Gamma_0)$ consists of  high peak close to zero, and
very long tail towards high values. This is because 
the density of states
\be
\rho(E)=\left\langle\sum_n \delta(E-E_n)\right\rangle \approx 
\sum_n \ds{\frac{\eps/2\pi}{(E-E_n)^2+\eps^2}}
\ee
consists of set of very narrow  separated peaks centered around eigenenergies $E_n$
for small $\epsilon$. Numerical data for $\Gamma_0$ become reliable
only for larger values of $\epsilon$, for which the density of states  is a smooth
function of the energy. As shown in Fig. \ref{distrx}, $P$
converges to the Gaussian distribution, independent on $\epsilon$  for sufficiently large 
values of $\epsilon$.

\subsection{Singularity of $\Gamma_0$ for $\omega\to 0$}

Figure  \ref{gamma-1} shows numerical data for the imaginary part of $\Gamma_0(E,\omega)$ 
as a function of $\omega$ for three  values of the energy
$E=5$, $E=6.58$ (the mobility edge) and $E=7$.  Data prove the singular behavior
$\Gamma_0\sim 1/\omega$. 
Comparison of numerical data calculated for two and more different values of $\epsilon$
enable us also to estimate the accuracy of our results. 
Although the singularity  $\sim 1/\omega$ transforms to
 Im~$\Gamma_0\sim \omega/(\omega^2+\epsilon^2)$ when $\eps\ne 0$.
Im~$\Gamma_0$ becomes independent on $\epsilon$ 
for $\omega\gg\epsilon$ 
(the region left from the vertical line in Fig. \ref{gamma-1}). 
Only these data were used for the calculation of the coefficient $B(E)$.

In Fig. \ref{gamma-2} we plot numerical data for the coefficient  $B(E)$
and compare them  with $2\pi\rho(E)$. 
The density of state $\rho(E)$ was calculated by direct diagonalizing of the 
Hamiltonian for $L=8$ and $L=16$. To increase the number of eigenstates,
 statistical ensembles of $N=10^3$ samples were used for each $L$. 
As shown in Fig. \ref{gamma-2}, the density of states in the band 
tail still depends on the system size.
Clearly, $L=8$ is not sufficient for the calculation of $\rho$. 
To check the convergence of the density of states,
we calculated the density of states for the energy $E=7$, also  from the statistical ensemble
of samples of the size $L=40$. The eigenenergies were calculated by Lanczos algorithm. 
\cite{BM} A comparison of the obtained density of states for $L=16$ and 40 
[$\rho(E=7)=0.0076$ for $L=16$,
and  $\rho(E=7)=0.0087$ for $L=40$] indicates that the convergence of the density of states
is very slow in the band tail. 

As shown in Fig. \ref{gamma-2},  obtained coefficient $B(E)$ agrees very well with our data
 for the density of states,
\be\label{tt}
B(E)=2\pi\rho(E).
\ee
The agreement is even better when we compare $\rho$ and $B(E)$  calculated for 
the same system size. 
Since the size corrections of both $B(E)$ and $\rho(E)$ are positive
(both quantities increases when $L$ increases), we conclude that our data
for $B(E)$ showed in Fig. \ref{gamma-2} definitely exclude the possibility that $B(E)$
decreases to zero when $E$ approaches the mobility edge.

\begin{figure}
\psfrag{a}{${\rm Re}~q^2\Gamma_q$}
\psfrag{b}{$q^2=1$}
\psfrag{c}{${\rm Im}~q^2\Gamma_q$}
\psfrag{d}{$q^2=2$}
\psfrag{e}{$q^2=3$}
\psfrag{g}{$q^2\Gamma_q$}
\psfrag{o}{$\omega/q^2$}
\includegraphics[clip,width=0.45\textwidth]{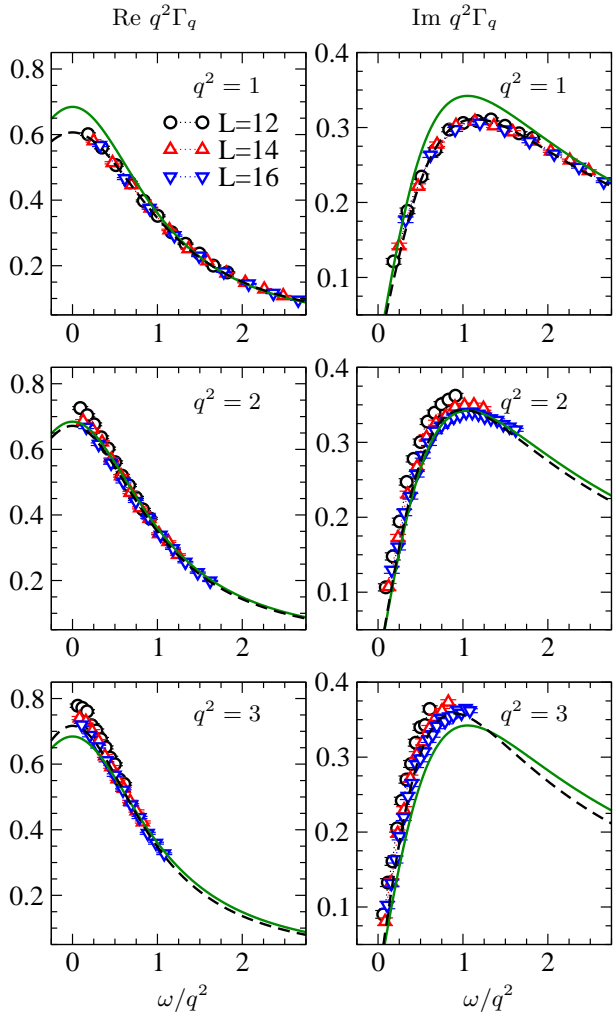}
\caption{%
The real and the imaginary part of $q^2\Gamma_q(E,\omega)$,
given by Eq. (\ref{g-kov1}), as a function of
$\omega/q^2$ for $E=0$ (the band center) and  for three values of $q$, 
$\vec{q}=(2\pi/L)[1,0,0]$, $(2\pi/L)[1,1,0]$ and $(2\pi/L)[1,1,1]$.
Solid lines show theoretical prediction, given by Eq. (\ref{g-kov1}).
with  $D=1.055$.
Dashed lines are fits of numerical data for $L=16$ with $0.94<D<1.15$.
}
\label{gamma-3}
\end{figure}

\begin{figure}
\psfrag{a}{${\rm Re}~q^2\Gamma_q$}
\psfrag{b}{$P({\rm Re}~q^2\Gamma_q)$}
\includegraphics[clip,width=0.45\textwidth]{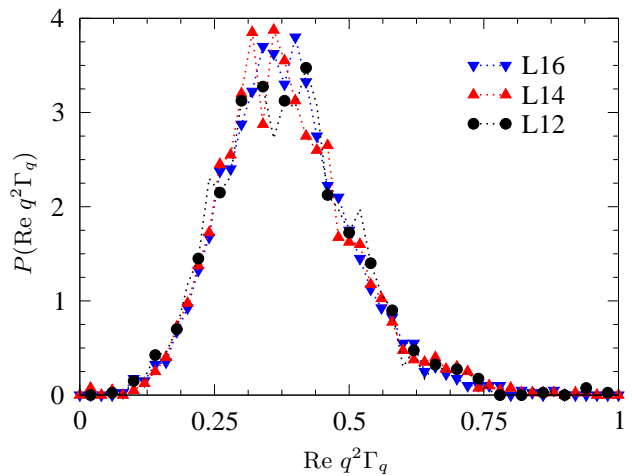}
\caption{(color online)
The probability distribution of $q^2\Gamma_q$ for $q=1$ and three system size, 
$L=12$, 14 and 16 and $\omega/q^2\approx 1$. The width of the distribution does not
depend on the size of the system.  The mean value, $\langle q^2\Gamma_q\rangle = 0.37$, 
and the variance, var $q^2\Gamma_q \approx 0.11$. does not depend on the size of the system.
}
\label{distr}
\end{figure}

\subsection{$\Gamma_q$ for non-zero $q$}

In this section we calculate the electron-hole correlation function $\Gamma_q$ for non-zero values of $q$.
We show that numerical data are consistent with theoretical prediction. From
numerical data,  we
estimate  the diffusion constant $D$.

In general, $D$ is a function of both $\omega$ and $q$. Numerical analysis of $D(\omega,q)$
for critical disorder $W=W_c$ and energy $E=0$ was performed in Ref. \cite{brandes}
Numerical simulations confirmed scaling behavior of the diffusive constant 
in the critical regime, predicted theoretically . \cite{chalker}
Since the critical region around the mobility edge $E=E_c$ is very narrow, \cite{BM}
 we restrict our analysis to the metallic regime, $W=2\ll W_c$ and $E=0$. Here, we
expect that $D$ is constant, independent on the frequency and wave vector.

Fig. \ref{gamma-3} shows the real and the imaginary parts of the function
$q^2\Gamma_q$, 
\be\label{g-kov1}
q^2\Gamma_q(E,\omega)=\ds{\frac{2\pi\rho(E)}{-i\omega/q^2+D}},
\ee
as a function of $\omega/q^2$.
The size of  the system is $L=12$, 14 and 16
with periodic boundary conditions. Three values of the wave vector were considered:
$\vec{q}$: $(1,0,0)$, $(1,1,0)$
and $(1,1,1)$ (in units of $2\pi/L$).  
As discussed above, $\epsilon=5/L^3$ is already sufficient for numerical analysis
of the electron-hole correlation function at the band center. We use this value in all calculations below. 

Numerical data lie on the one universal curve. This universality is better pronounced
for small values of $q$. Stronger finite size effects are observed for larger $q$.
Data for $L=16$ are fitted to Eq. (\ref{g-kov1}) with fixed density of states,
$2\pi\rho(E=0)=0.72$ and free parameter $D$. Fits are shown in Fig. \ref{gamma-3} by dashed lines.
From fits, we  estimate the  diffusion constant,
\be\label{d}
D\approx 1.05\pm 0.10.
\ee 
This value is compared with diffusion constant calculated 
by the transfer matrix method from  
the $L-$ dependence of the conductance, 
\be\label{gg}
g(L)=\sigma L 
\ee
where $\sigma=e^2D(E)\rho(E)$ is the conductivity.  \cite{acta}
We obtained $D\approx 1.055$,  which perfectly agrees with our estimation (\ref{d}).

Finally, we present in Fig. \ref{distr} the 
probability distribution of real part of $q^2 \Gamma_q$ 
for three sizes of the system  $\omega/q^2\approx 1$ for all systems.  
Our results confirm that the distribution
 depends only  on the ratio $\omega/q^2$.  This is consistent with Eq. (\ref{g-kov1}). 
More important, data in Fig. \ref{distr} indicate that $P$ does not depend on the system size.

\section{Inverse participation ratio}

\begin{figure}[b!]
\psfrag{x}{$L$}
\psfrag{y}{$Y$}
\psfrag{j}{$\ln({\rm I})-a_{E}\ln(L)$}
\psfrag{k}{$P(\ln I)$}
\includegraphics[clip,width=0.45\textwidth]{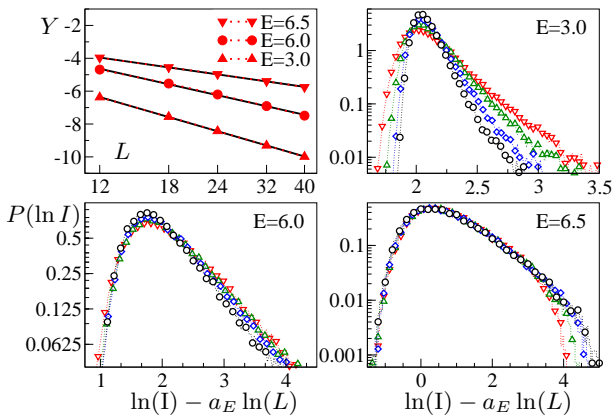}
\caption{(Color online) The left upper panel shows 
the system size dependence of $Y\propto  a_E \ln L$ with $a_3=-3.006$,
$a_6=-2.326$ and $a_{6.5}=-1.48$. Three other panels show the probability 
distribution $P(\ln I)$ for three energies of the electron and for the system size $L=18$
($\bigtriangledown$), 24
($\bigtriangleup$), 
32 ($\diamond$)  
and $L=40$ ($\circ$).
 Note the scaling of the horizontal axis.
$N_{\rm stat}=10^5$  eigenstates from the interval  $|E_n-E|<0.025$ were used for 
the statistics. 
 \cite{BM}}
\label{bm-fig1}
\end{figure}

The absence of localized states in the metallic phase can be 
demonstrated also by the analysis of the probability distribution of
inverse participation ratio,
defined as
 \cite{Kramer,EM}
\be\label{ipr}
I(E_n)=\sum_{r} |\Phi_n(r)|^{4}.
\ee
Here, $E_n$ and $\Phi_n(r)$  is the $n$th eigenenergy  and eigenfunction of the Hamiltonian (\ref{ham}),
respectively. 
If the $|E|<E_c$, then we expect that all  eigenstates are  conductive, 
with the  wave functions distributed moreless homogeneously throughout the sample, so that
$|\Phi_n(r)|^2\propto L^{-d}$. Inserting in Eq. (\ref{ipr}) we obtain that 
$I(E_n)\propto L^{-3}$(in the 3D system).
On the other hand, the wave function of  localized electron  is non-zero only 
in a small region, 
where $|\Phi_n(r)|\sim 1$. Hence, 
 $I(E_n)\sim 1$.   The size dependence of $I(E_n)$ in the critical region
 deserves more detailed analysis since the spatial distribution of electron is 
 multifractal
 \cite{EM,evers2000} and $I\propto L^{-d_2}$ where $d_2\approx 1.28$. \cite{BM}

\subsection{The size dependence of IPR}

The energy spectrum of the Hamiltonian depends on the system size, $L$, and on the microscopic
details of the disorder in a given sample.  For a given  system 
size, we consider a statistical ensemble of $N_s$ samples which differ only in the realization
of random energies, $\epsilon_{\vec{r}}$. For each sample, 
we calculate   all eigenenergies, $E_n$, lying  in a narrow
energy interval, $E-\delta,E+\delta$, and calculate corresponding $I\equiv I(E_n)$. 
For the $i$th sample, the number of eigenstates, 
$n_i$, depends on the microscopic realization of the disorder. 

Collecting  $N_{\rm stat}=\sum_i n_i$ values of IPR, we can construct
its probability distribution $P(I)$ or $P(\ln I)$.
Since the values of $I$ might fluctuate in many orders in magnitude
in the critical region \cite{mirlin},
it is  more convenient to use  the  logarithm of $I$ and the mean value,
\be\label{Y}
Y\equiv Y(E)=\displaystyle{\frac{1}{N_{\rm stat}}}\sum_i^{N_s}\sum_{|E-E_n|<\delta}\ln I(E_n).
\ee

\begin{figure}[t]
\psfrag{y}{~~~~~~~~~~$\Pi_{\im}$}
\includegraphics[clip,width=0.45\textwidth]{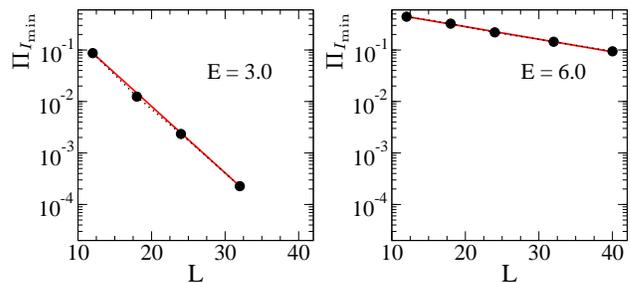}
\caption{(Color online) %
Left panel: The size dependence of the probability $\Pi$, defined by Eq. 
(\ref{chvost}) with  $I_{\rm min}=-2\ln L$ for
the energy $E=3$.  
The right panel shows the probability $\Pi_{\im}$ for  $I_{\rm min}=-3/2\ln L$ and $E=6$.
Data confirm that $\Pi_{\im}$ decreases exponentially when the size of the system increases.
}
\label{pokles}
\end{figure}

The upper left panel of Fig.  \ref{bm-fig1} shows 
the system size dependence of $Y$  for three energies below the mobility edge. 
We found that $Y\sim a_E \ln L$.
In the metallic regime, we expect $a_E=-3$ for all $|E|<E_c$.
At the mobility edge,  $a_{E=E_c}= d_2$, the fractal dimension \cite{BM}.
For $E=3$, which is the energy far below the mobility edge, 
we indeed find $a_3=-3$, in agreement with our expectations.
Higher values of $a_E$ obtained for energies closer to the mobility edge 
are due to the finite size effects. We expect that $a_E$ converges to $ -3$ when 
the size of the system increases, $L\to\infty$.
This is consistent with analytical expression for the mean IPR, \cite{fyod} 
\be\label{fyo}
\langle I(L)\rangle =L^{-3}[1+4L/(\ell g)]. 
\ee
Deep in the metallic regime, $|E|<E_c$, the conductance $g$
is $\propto L$ (Eq. (\ref{gg}), so that Eq. (\ref{fyo})
 reproduces $\langle I\rangle\sim L^{-3}$.
 However, 
the linear increase of $g \sim L$ can be obtained only when the size of the system
 $L\gg \xi$. For smaller-size system, $L\sim \xi$, the correction term $4L/\ell g$ in Eq. (\ref{fyo})
becomes $L$ dependent
and causes the deviation from the $L^{-3}$-\ dependence of mean $I$.
The scaling behavior of IPR for energies close to the mobility edge is discussed later
in section \ref{FSE}. 

\subsection{Probability distribution of IPR}

Because of the randomness and wave character of the electron motion,
the mean value of any quantity might not provide us with the entire information about the
system. For instance, although $Y\propto-3\ln L$, there still might exists
some localized electronic states with an eigenvalue $E_n$ and  $\ln I\sim 0$.
To measure the probability that insulating states exist in the metallic phase 
($|E|<E_c$),  we  plot in Fig. \ref{bm-fig1}
the probability distribution $P(\ln I)$
calculated for three  energies and various size of the system.  

Data for $E=3$ confirm that 
$P(\ln I)$ gets narrower when $L$ increases.  
This is consistent with analytical result, var~$I\sim L^{-2}$. \cite{fyod}
The narrowing of the probability distribution  is less visible for energies close
to the mobility edge $E_c$. 
\begin{figure}[t]
\psfrag{x}{\large$\ln I$}
\psfrag{y}{\large${\rm P}(\ln I)$}
\includegraphics[clip,width=0.4\textwidth]{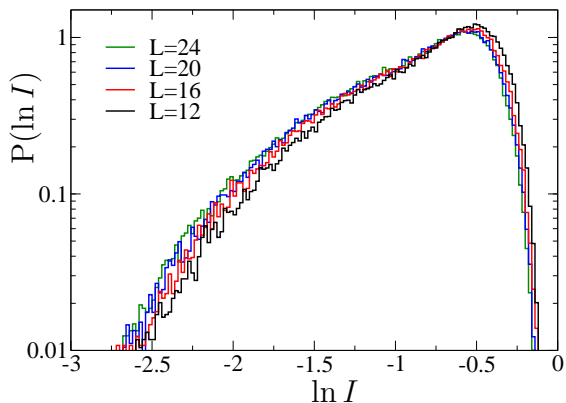}
\caption{(Color online) %
Probability distribution $P(\ln I)$ for the localized states 
for $L= 12$, 16, 20 and 24.
The distribution only weakly  depends on the size of the system
and decreases exponentially when $\ln I$ decreases.
Consequently, the probability to find, for instance, the eigenstate with $\ln I< -\ln L$
decreases exponentially when $L$ increases. Therefore, 
the probability to find the conductive state 
in the localized phase ($|E|>E_c$) is exponentially small. 
}
\label{IPR-2}
\end{figure}

Since the mean value, $Y$, decreases as $\sim a_E\ln L$ when
$L$ increases, the existence of localized states is possible only if $P(\ln I)$ possesses
 a long tail which  assures a non-zero probability to have $\ln I\sim 0$ for any system size. 
However, our data in Fig. \ref{bm-fig1} show that this is not the case.
Contrary,  $P(\ln I)$ decreases exponentially for $\ln I$ larger than $Y$.
To measure this exponential decrease quantitatively,  we 
 calculate 
the probability that  $\ln I$ is larger than certain value, $\im$: 
\be\label{chvost}
\Pi_{\im} = \int_{\im}^\infty {\rm d} \ln I' P(\ln I') = \int_{\im}^\infty {\rm d} I' P(I')
\ee
We choose $\im=L^{-2}$ for $E=3$ and $\im=L^{-3/2}$ for $E=6.0$.
In Fig. \ref{pokles} we prove that  $\Pi_{\im}$ decreases exponentially 
as a function of the size of the system $L$. This is consistent with theoretical prediction
$P(I)\sim\exp(-\alpha I)$. \cite{mirlin} Note, this exponential decrease is visible already for energies very close to the critical point,
(right panel of 
Fig. \ref{pokles}).
Since $I(E_n)\sim 1$ for the localized state $E_n$, 
the probability to observe
the localized state inside  the metallic phase  decreases exponentially 
when the size of the system increases. We conclude that the probability to find any localized state
is zero in the limit of $L\to\infty$.

Similar conclusion, namely that there are no metallic states in the energy interval 
$E>E_c$, can be derived  for localized phase. In Fig. \ref{IPR-2} we show the probability distribution $P(\ln I)$ for eigenstates around the energy $E=7.5$. Clearly, the distribution is
size independent, and decreases exponentially for small values of $\ln I$. Since metallic state requires
that $\ln I\sim -3\ln L$, we conclude that there are no metallic states in the insulating phase. 

\begin{figure}[t]
\psfrag{x}{$\ln I - d_{2}\ln(L)$}
\psfrag{y}{$P(\ln I)$}
\includegraphics[clip,width=0.35\textheight]{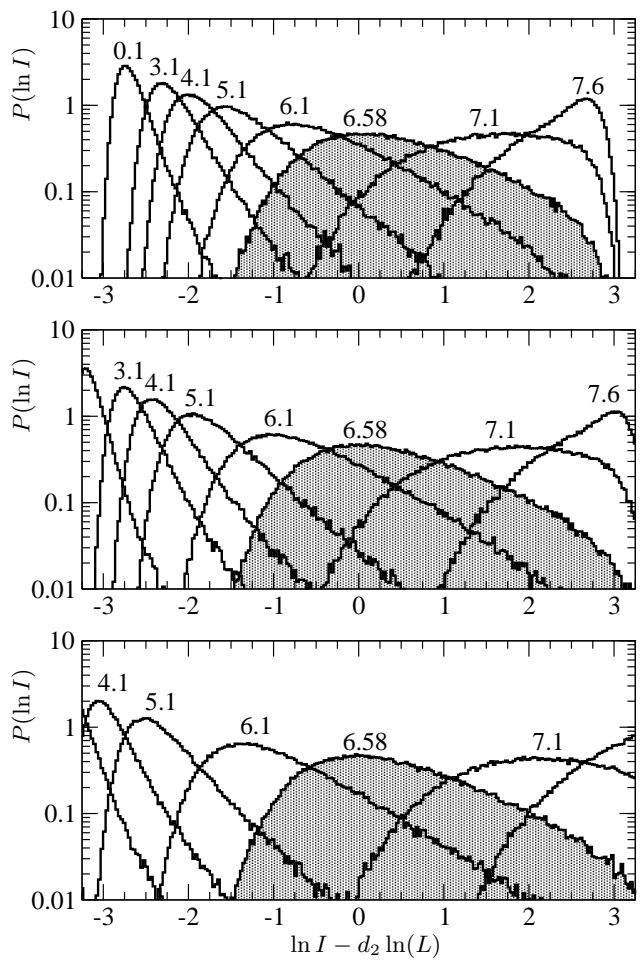}
\caption{
Probability distribution $P(\ln I)$ for various energies. 
and three sizes of the system: $L=12$ (top), $L=16$ (middle) and
$L=24$ (bottom). Shaded area is the critical distribution $P_c$ for $E=E_c$.
Note, horizontal axis is scaled by $-d_2\ln L$ in order to keep the critical
distribution for $E_c=6.58$ in the center of  the figure. 
}
\label{IPR-3}
\end{figure}

\subsection{Finite size scaling}\label{FSE}

The right upper panel of Fig. \ref{bm-fig1}  shows the distribution of IPR in the
metallic regime. 
The distribution is centered at $3\ln L$, in agreement with our expectation.
More important is the form of the distribution for larger values of $\ln I$. Our data show
that the probability to observe $\ln I\sim -2\ln L$ decreases exponentially when the size of
the system increases. Hence, we conclude that there are no localized states.

However, as is shown in lower panels of Fig. \ref{bm-fig1},
narrowing of the distribution  $P(\ln I)$ can be numerically observed only when 
the energy $E$ lies deep in the metallic phase.  This is 
easy to understood. 
The metallic phase can be observed only in systems of size $L\gg \xi(E)$,
where $\xi(E)$ is the correlation length. \cite{MacKK} Since $\xi(E)$ diverges
as $\xi(E)\propto |E_c-E|^{-\nu}$ ($\nu\approx 1.57$ is the critical exponent)
when  $E\to E_c^-$, we  cannot observe 
the metallic behavior 
for energies close to the mobility edge and for fixed size of the system.
Nevertheless,
reliable conclusion about the character of metallic states in the vicinity of the mobility edge 
can be drawn with the use of the finite size scaling analysis.
\cite{AALR}
The probability distribution $P(\ln I)$ depends 
both on the energy $E$ and on the size of 
the system $L$.
Following the  scaling theory,  $P$ calculated for a 
given energy $E$ and size $L$ is equivalent 
to $P$ obtained for $E'$ closer to $E_c$, but larger system size $L'>L$.

Scaling of IPR in the critical region  was numerically proved in Ref. \cite{BM}.  
In Fig. \ref{IPR-3} we demonstrate how the
scaling idea works in the metallic phase.
The distributions $P(\ln I)$ were calculated for various energies of the electron 
and for three size of the system.  We   see
the similarity in the form of $P(\ln I)$ calculated for different $E$ and $L$.
For instance, the distribution for $E=3.1$ and $L=16$
is similar  to the distribution for $E=0.1$ and $L=12$. Similarly,
we can compare $E=5.1$ and  $L=24$ with that for  $E=4.1$, $L=12$. 
We also see that the  form of the probability distribution
$P(\ln I)$ only weakly depends on the energy $E$ of the electron 
when  $L\gg \xi(E)$.
From this similarity we conclude that the properties of the distribution $P(\ln I)$ 
are universal in the metallic phase when $L\to\infty$. Therefore,
the probability to find the localized state decreases to zero for any energy
$|E|<E_c$.

\section{Conclusion}

We studied numerically  two parameters, important for the construction of 
the analytical theory of the metal-insulator transition.  
First, 
we verified the relation between electron-hole correlation 
function $\Gamma$ and the density of states,
Eq. (\ref{t}). We proved that this relation  not only holds for all energies of the electron,
both in the metallic and localized phase, but can be recovered for any size of the system. 
Our numerical procedure enables us to calculate, from $\Gamma$, the diffusion constant $D$.
In the metallic limit, $D$ agrees with estimation from the transfer matrix.
Also, numerical data  indicate that $\Gamma_q(E,\omega)$ is not the self-averaged quantity 
in the metallic regime.  

It would be interesting 
to investigate also the scaling behavior of the diffusive constant
in the critical regime. \cite{chalker}
Such analysis  could confirm numerical scaling observed recently 
for the case of the critical disorder at band center. \cite{brandes} 
However, since
only frequencies  $\omega>\eps$ are relevant  in numerical data,  we have to analyze much
larger system size in order to fit both energies $E\pm\omega/2$ into the narrow 
critical region. \cite{BM}

We also present numerical data for the 
mean values and for probability distributions  of the inverse participation ratio.
Our data, consistent with previous analytical results,
enable us to prove that there are no
localized states inside the metallic phase. 
All electron states are extended, and the probability
to find  the state which does not span through the sample decreases exponentially 
to zero when the size
of the system increases. Similarly, no metallic states were observed 
on the opposite side of the 
critical point: in the insulating tail of the spectra, all electronic states are localized.
Although this result seems to be easily accepted \cite{Kramer}, 
it was never proved numerically.

In contrast to the analytical theory, numerical methods do not enable the
analysis  of the behavior of any quantity 
in the limit of infinite system size. We can only describe how the  
variables of interest change 
when the size of the system increases. With the use of the finite size scaling hypothesis, 
we conclude that our results remain valid also in the limit of $L\to\infty$. Since all data were
obtained  without any additional assumption about the averaging procedure or the statistics 
of parameters of interest, 
they can serve as a starting point for the construction of the
analytical theory of the Anderson transition.

\medskip

This work was supported by grant APVV, project n. 51-003505 and VEGA, project n. 2/6069/26.

\end{document}